\documentclass[twocolumn,showpacs,aps,pre]{revtex4-1}

\usepackage{amssymb,amsmath}
\usepackage[dvips]{graphicx}

\begin{document}

\title{Enhanced rare region effects in the contact process with long-range correlated disorder}

\author{Ahmed K. Ibrahim}
\affiliation{Department of Physics, Missouri University of Science and Technology,
Rolla, MO 65409, USA}

\author{Hatem Barghathi}
\affiliation{Department of Physics, Missouri University of Science and Technology,
Rolla, MO 65409, USA}

\author{Thomas Vojta}
\affiliation{Department of Physics, Missouri University of Science and Technology,
Rolla, MO 65409, USA}

\begin{abstract}
We investigate the nonequilibrium phase transition in the disordered
contact process in the
presence of long-range spatial disorder correlations. These correlations greatly increase the probability for
finding rare regions that are locally in the active phase while the bulk system is still in the inactive phase.
Specifically, if the correlations decay as a power of the distance, the rare region probability
is a stretched exponential of the rare region size rather than a simple exponential as is the case for uncorrelated
disorder. As a result, the Griffiths singularities are enhanced and take a non-power-law form.
The critical point itself is of infinite-randomness type but with critical exponent values that differ from the uncorrelated
case. We report large-scale Monte-Carlo simulations that verify and illustrate our theory.
We also discuss generalizations to higher dimensions and applications to other systems such as the
random transverse-field Ising model, itinerant magnets and the superconductor-metal transition.
\end{abstract}
\date{\today}

\maketitle


\section{Introduction}
\label{sec:Intro}

The effects of quenched spatial disorder on phase transitions have been a topic of
great interest for several decades. Initially, research concentrated on classical
(thermal) transitions for which many results can be obtained by using perturbative
methods adapted from the theory of phase transitions in clean systems
(see, e.g., Ref.\ \cite{Grinstein85}).

Later, it became clear, however, that many
transitions are dominated by the non-perturbative effects
of strong, rare disorder fluctuations and the rare spatial regions that support them.
Such rare regions can be locally in one phase while the bulk system is in the other.
The resulting slow dynamics leads to thermodynamic singularities, now known as the
Griffiths singularities \cite{Griffiths69,McCoy69}, not just at the transition point
but in an entire parameter region around it. Griffiths singularities at generic
classical (thermal) phase transitions are very weak and probably unobservable
in experiment \cite{Imry77}. In contrast, at many quantum and
nonequilibrium phase transitions, the rare regions lead to strong Griffiths effects
characterized by non-universal power-law singularities of various observables.
The critical point itself is of exotic infinite-randomness type and characterized
by activated rather than power-law dynamical scaling.
This was first demonstrated in the random-transverse field Ising chain using
a strong-disorder renormalization group \cite{Fisher92,Fisher95} as well as
heuristic optimal fluctuation arguments and computer simulations \cite{ThillHuse95,YoungRieger96}
\footnote{Partial results on the related McCoy-Wu model had already been obtained much earlier
\cite{McCoyWu68,McCoyWu68a} but they were only fully understood after Fisher's
strong-disorder renormalization group calculation \cite{Fisher92,Fisher95}}.
Similar power-law Griffiths singularities were also found at the nonequilibrium
transition of the disordered contact process \cite{Noest86,*Noest88,HooyberghsIgloiVanderzande03,*HooyberghsIgloiVanderzande04,VojtaDickison05}
and at many other quantum and nonequilibrium transitions.  In some systems,
the rare region effects are even stronger and destroy the sharp phase transition
by smearing \cite{Vojta03a,*Vojta03b,*Vojta04}.
Recent reviews and a classification of rare region effects can be found, e.g., in Refs.\
\cite{Vojta06,*Vojta10}.

The majority of the literature on rare regions and Griffiths singularities
focuses on uncorrelated disorder. In many physical situations, we can expect, however,
that the disorder is correlated in space, for example if it caused by charged
impurities. It is intuitively clear that sufficiently
long-ranged spatial disorder correlations must enhance the rare region effects
because they greatly increase the probability for finding large atypical rare regions.
Rieger and Igloi \cite{RiegerIgloi99} studied a random transverse-field
Ising chain with power-law disorder correlations. They indeed found that
sufficiently long-ranged correlations change the universality class of the
transition. They also predicted that the Griffiths singularities take the same
power-law form as in the case of uncorrelated disorder, but with changed exponents.

In this paper, we investigate the nonequilibrium phase transition in the
disordered one-dimensional contact process with power-law disorder correlations
by means of optimal fluctuation theory and computer simulations. Our paper is
organized as follows. We define the contact process with correlated
disorder in Sec.\ \ref{sec:Model}.  In Sec.\ \ref{sec:Theory}, we develop our theory
of the nonequilibrium phase transition and the accompanying Griffiths phase.
Specifically, we show that the probability of finding a large rare region is a
stretched exponential of its size rather than a simple exponential as for uncorrelated
disorder. As a result, the Griffiths singularities are enhanced and take a non-power-law form.
The critical point itself is of infinite-randomness type but its exponents differ from
the uncorrelated case. Section \ref{sec:MC} is devoted to Monte-Carlo simulations
that verify and illustrate our theory. In Sec.\ \ref{sec:General}, we generalize our results
to higher dimensions and other physical systems. We also discuss the relation between the present
work and Ref.\ \cite{RiegerIgloi99}. We conclude in Sec.\ \ref{sec:Conclusions}.

\section{Contact process with correlated disorder}
\label{sec:Model}

The contact process \cite{HarrisTE74} is a prototypical nonequilibrium many-particle
system which can be understood as a model for the spreading of an epidemic.
Consider a one-dimensional regular lattice of $L$ sites. Each site can be in one of two
states, either inactive (healthy) or active (infected). The time evolution of the contact
process is given by a continuous-time Markov process during which active lattice sites infect
their nearest neighbors or heal spontaneously. Specifically, an active site becomes
inactive at rate $\mu$, while an inactive site becomes active at rate $n\lambda/2$ where
$n$ is the number of its active nearest neighbors. The healing rate $\mu$ and the infection
rate $\lambda$ are the external control parameters of the contact process.
Without loss of generality, $\mu$ can be set to unity, thereby fixing the unit of time.

The qualitative behavior of the contact process is easily understood. If healing
dominates over infection, $\mu \gg \lambda$, the epidemic eventually dies out completely,
i.e., all lattices sites become inactive. At this point, the system is in a fluctuationless
state that it can never leave. This absorbing state constitutes the inactive phase of
the contact process. In the opposite limit, $\mu \ll \lambda$, the infection never
dies out (in the thermodynamic limit $L\to \infty$). The system eventually reaches
a steady state in which a nonzero fraction of lattices sites is active.
This fluctuating steady state constitutes the active phase of the contact process.
The active and inactive phases are separated by a nonequilibrium phase transition
in the directed percolation universality class \cite{GrassbergerdelaTorre79, Janssen81,Grassberger82}.
The order parameter of this absorbing-state transition is given by the steady
state density $\rho_{\rm stat} = \lim_{t\to\infty} \rho(t)$ which is the long-time
limit of the density of infected sites at time $t$,
\begin{equation}
\rho(t) = \frac 1 {L} \sum_{i} \langle  n_{i}(t) \rangle~.
\label{eq:rho_definition}
\end{equation}
Here, $n_{i}(t)$ is the occupation of site ${i}$ at time $t$, i.e.,
$n_i(t)=1$ if the site is infected and $n_i(t)=0$
if it is healthy.  $\langle \ldots \rangle$ denotes the
average over all realizations of the Markov process.

So far, we have discussed the clean contact process for which $\lambda$ and $\mu$ are spatially uniform.
Quenched spatial disorder is introduced by making the infection rate $\lambda_i$ of site $i$ and/or its
healing rate $\mu_i$ random variables. The correlations of the randomness can be characterized by
the correlation function
\begin{equation}
G_\lambda(i,j) = [\lambda_i \lambda_j]_\textrm{dis} - [\lambda_i]_\textrm{dis} \, [\lambda_j]_\textrm{dis}
\label{eq:Gij}
\end{equation}
where $[\ldots]_\textrm{dis}$ denotes the disorder average. The correlation function $G_\mu$ of
the healing rates $\mu_i$ can be defined analogously.
The existing literature on the disordered contact process mostly considered the case of
uncorrelated disorder, $G_\lambda(i,j) \sim G_\mu(i,j) \sim \delta_{ij}$.
In the present paper, we are interested in long-range correlations whose correlation function
decays as a power of the distance $r_{ij}$ between the two sites,
\begin{equation}
G_\lambda(i,j) \sim G_\mu(i,j) \sim r_{ij}^{-\gamma}~,
\label{eq:G_power}
\end{equation}
for large $r_{ij}$.
For our analytical calculations we will often use a correlated Gaussian distribution
\begin{equation}
P_G(\lambda_1, \dots, \lambda_L) \sim \exp \left[-\frac 1 2 \sum_{i,j} (\lambda_i-\bar\lambda) A_{ij} (\lambda_j-\bar\lambda ) \right]
\label{eq:Gaussian}
\end{equation}
of average $\bar\lambda=[\lambda_i]_\textrm{dis}$ and covariance matrix $(A^{-1})_{ij} = G_\lambda(i,j)$
\footnote{The negative tail of the Gaussian has to be truncated appropriately because the infection rate $\lambda_i$
must be positive.}. Alternatively, we will also use a correlated binary distribution in which
$\lambda_i$ can take values $\lambda$ and $c\lambda$ with overall probabilities $(1-p)$ and $p$, respectively.
Here, $p$ and $c$ are constants between 0 and 1.

\section{Theory}
\label{sec:Theory}
\subsection{Rare region probability}
\label{subsec:Theory_RR}

The Griffiths phase in the disordered contact process is caused by rare large spatial regions
whose effective infection rate  is larger than the bulk average $\bar\lambda$.
For weak disorder and outside the asymptotic critical region, the effective infection rate
can be approximated by
\begin{equation}
\lambda_{RR} \approx \frac 1 {L_{RR}} \sum_{i \in RR} \lambda_i
\label{eq:lambda_RR}
\end{equation}
To estimate how the probability distribution of $\lambda_{RR}$ depends on the rare region size $L_{RR}$,
we start from the correlated Gaussian (\ref{eq:Gaussian}), introduce
$\lambda_{RR}$ as a new variable and then integrate out all other random variables.
For large $L_{RR}$ and up to subleading boundary terms, this leads to the distribution
\begin{equation}
P(\lambda_{RR},L_{RR}) \sim \exp \left[ -\frac {L_{RR}} {2 \tilde G(L_{RR})} (\lambda_{RR}-\bar\lambda)^2      \right]
\label{eq:PlambdaRR}
\end{equation}
where $\tilde G(L_{RR})$ is the sum over the correlation function
\begin{equation}
\tilde G(L_{RR}) \sim \sum_{j=0}^{L_{RR}/2} G_{\lambda}(0,j)~.
\label{eq:tilde_G}
\end{equation}
Two cases need to be distinguished, depending on the value of the decay exponent $\gamma$
in the correlation function (\ref{eq:G_power}).
If $\gamma >1$, the sum $\tilde G(L_{RR})$ converges in the limit $L_{RR} \to \infty$.
The probability distribution of the effective infection rate $\lambda_{RR}$ thus takes
the asymptotic form
\begin{equation}
P(\lambda_{RR},L_{RR}) \sim \exp \left[ - \frac 1 {2b^2} \, {L_{RR}}\, (\lambda_{RR}-\bar\lambda)^2  \right]
\label{eq:PlambdaRR-sr}
\end{equation}
where $b$ is a constant. This form is identical to the result for uncorrelated or short-range correlated disorder
(and agrees with the prediction of the central limit theorem).
For $0< \gamma < 1$, in contrast, the sum $\tilde G(L_{RR})$ behaves as
$L_{RR}^{1-\gamma}$  for large $L_{RR}$. Consequently, the probability distribution
of $\lambda_{RR}$ reads
\begin{equation}
P(\lambda_{RR},L_{RR}) \sim \exp \left[ - \frac 1 {2b^2} \, {L_{RR}^\gamma}\, (\lambda_{RR}-\bar\lambda)^2  \right]~.
\label{eq:PlambdaRR-lr}
\end{equation}
This is a stretched exponential decay in $L_{RR}$ rather than the simple exponential obtained in
(\ref{eq:PlambdaRR-sr}). In other words, for $0< \gamma < 1$, the probability for finding a large
deviation of $\lambda_{RR}$ from the average $\bar\lambda$ decays much more slowly with rare region size
than in the uncorrelated case.

We have also considered a correlated binary disorder distribution instead of the Gaussian (\ref{eq:Gaussian}).
In this case, rare regions can be defined as regions of $L_{RR}$ consecutive sites having the larger
of the two infection rates. For uncorrelated disorder, the probability for finding such a region decays
as a simple exponential of its size $L_{RR}$. We have confirmed numerically that the corresponding probability
for the power-law correlations (\ref{eq:G_power}) with $0<\gamma<1$ follows a stretched exponential
\begin{equation}
w(L_{RR}) \sim \exp(-c L_{RR}^\gamma)
\label{eq:w(L_RR)}
\end{equation}
with the same exponent $\gamma$ as in eq.\ (\ref{eq:PlambdaRR-lr}).

\subsection{Griffiths phase}
\label{subsec:Theory_Griffiths}

We now use the results of Sec.\ \ref{subsec:Theory_RR} to analyze the time evolution
of the density of active sites $\rho(t)$ in the Griffiths phase on the inactive side of
the nonequilibrium transition. This calculation is a generalization to the case of correlated disorder
of the approach of Refs.\ \cite{VojtaHoyos14,VojtaIgoHoyos14}.

The rare region contribution to $\rho(t)$ can be obtained by summing over all regions
that are locally in the active phase, ie., all regions having $\lambda_{RR}>\lambda_c$.
For the correlated Gaussian distribution (\ref{eq:Gaussian}), $\rho(t)$
reads
\begin{eqnarray}
\rho(t) &\sim& \int_{\lambda_c}^\infty d\lambda_{RR} \int_0^\infty dL_{RR}\,  P(\lambda_{RR},L_{RR}) \times \nonumber \\
&& \qquad \times L_{RR} \exp[-t/\tau(\lambda_{RR},L_{RR})]
\label{eq:rho_integral}
\end{eqnarray}
Here, $P(\lambda_{RR},L_{RR})$ is the rare region distribution (\ref{eq:PlambdaRR-sr}) or (\ref{eq:PlambdaRR-lr}),
depending on the value of $\gamma$; and $\tau(\lambda_{RR},L_{RR})$ denotes the lifetime of the rare
region. It can be estimated as follows.
As the rare region is locally in the active phase, $\lambda_{RR}>\lambda_c$, it can only decay via
an atypical coherent fluctuation of all its sites. The probability for this to happen
is exponentially small in the rare region size \cite{Noest86,*Noest88}, resulting
in an exponentially large life time
\begin{equation}
\tau(\lambda_{RR},L_{RR}) =  t_0 \exp\left[a L_{RR} \right]
\label{eq:tau}
\end{equation}
where $t_0$ is a microscopic time scale. The coefficient $a$ vanishes at $\lambda_{RR}=\lambda_c$
and increases with increasing $\lambda_{RR}$, i.e., the deeper the region is in the active phase,
the larger $a$ becomes. Because $a$ has the dimension of an inverse length, it scales as $\xi_\perp^{-1}$
(where $\xi_\perp$ is the correlation length)
according to finite-size scaling \cite{Barber_review83},
\begin{equation}
a = a' (\lambda_{RR}-\lambda_c)^{\nu_{0\perp}}~.
\label{eq:a_below}
\end{equation}
Note that $\nu_{0\perp}$ is the
\emph{clean} correlation length exponent unless the rare region is very close to
criticality (inside the narrow asymptotic critical region)
\footnote{The scaling behavior of $a$  also follows from the fact that the term $a L_{RR}$ in
the exponent of (\ref{eq:tau}) represents the number $(L_{RR}/\xi_\perp)$ of independent correlation volumes
that need to decay coherently.}.

In the long-time limit $t \gg t_0$, the integral (\ref{eq:rho_integral}) can be solved in
saddle-point approximation. The saddle point equations read
\begin{eqnarray}
\frac\partial {\partial L_{RR}} \left[\frac {L_{RR}^\gamma}{2b^2}(\lambda_{RR}-\bar\lambda)^2 + \frac {t}{t_0}e^{- a'(\lambda_{RR}-\lambda_c)^{\nu_{0\perp}} L_{RR}} \right]&=&0~,~~~
\label{eq:rho_sp_eq1}\\
\frac\partial {\partial \lambda_{RR}} \left[ \frac {L_{RR}^\gamma}{2b^2}(\lambda_{RR}-\bar\lambda)^2 + \frac {t}{t_0}e^{- a'(\lambda_{RR}-\lambda_c)^{\nu_{0\perp}} L_{RR}} \right]&=&0~,~~~
\label{eq:rho_sp_eq2}
\end{eqnarray}
and yield the saddle point values
\begin{eqnarray}
\lambda_{sp}-\lambda_c &=& \frac {\gamma\nu_{0\perp}}{2-\gamma\nu_{0\perp}} (\lambda_c-\bar\lambda)~,~~
\label{eq:lambda_sp}\\
L_{sp} &\sim& (\lambda_c-\bar\lambda)^{-\nu_{0\perp}} \, \ln(t/t_0)~.
\label{eq:L_sp}
\end{eqnarray}
Equations (\ref{eq:rho_sp_eq1}) to (\ref{eq:L_sp}) apply to the long-range correlated case $\gamma < 1$;
the corresponding relations for the short-range correlated case follow by formally setting $\gamma = 1$.

For the method to be valid, $\lambda_{sp}$ must be within the integration range of the
integral (\ref{eq:rho_integral}). The bulk system is in the inactive phase implying
$\bar\lambda < \lambda_c$. Moreover, the clean correlation length exponent of the
one-dimensional contact process takes the value $\nu_{0\perp} \approx 1.097$
\cite{Jensen99}. Consequently, the saddle-point value $\lambda_{sp}$ is larger
than $\lambda_c$, as required.
Inserting the saddle-point values into the integrand yields
\begin{equation}
\rho(t) \sim \exp\left[ -\frac 1 {z'} \left(\ln \frac t{t_0} \right)^\gamma \right]
\label{eq:rho_Griffiths}
\end{equation}
where
\begin{equation}
z' \sim (\lambda_c - \bar\lambda)^{\gamma\nu_{0\perp}-2}
\label{eq:z'}
\end{equation}
plays the role of a dynamical exponent in the Griffiths phase.
In the short-range correlated case, $\gamma$ is formally 1. Thus, eq.\ (\ref{eq:rho_Griffiths})
reproduces the well-known power-law Griffiths singularity of density in this case
\cite{Noest86,*Noest88,HooyberghsIgloiVanderzande03,VojtaDickison05}. In contrast,
in the long-range correlated case, $\gamma < 1$, the decay of the density is slower
than any power. Long-range disorder correlations thus lead to a qualitatively enhanced
Griffiths singularity.

The above derivation started from the correlated Gaussian distribution (\ref{eq:Gaussian}).
However, an analogous calculation can be performed for a correlated binary distribution by combining
the rare region probability (\ref{eq:w(L_RR)}) with the rare region life time (\ref{eq:tau}).
Solving the resulting integral over $L_{RR}$ in saddle-point approximation
leads to the same functional form (\ref{eq:rho_Griffiths}) of the Griffiths singularity,
with
\begin{equation}
z'=a^\gamma / c~.
\label{eq:z'_binary}
\end{equation}
If the rare regions are not in the active phase but right a the critical point,
their decay time depends on their size via the power law $\tau(\lambda_c,L_{RR}) \sim
L_{RR}^{z_0}$ rather than the exponential (\ref{eq:tau}). Here, $z_0\approx 1.581$ is the clean dynamical
exponent. For a correlated binary disorder distribution, this can be achieved by tuning the
stronger of the two infection rates to the clean critical value.
Repeating the saddle-point integration for this case gives a stretched
exponential density decay
\begin{equation}
\ln \rho(t) \sim -t^{\gamma/(\gamma +z_0)}~.
\label{eq:rho_clean_critical}
\end{equation}
As before, the short-range correlated case is recovered by formally setting $\gamma=1$.

Griffiths singularities in other quantities can be derived in an analogous manner. Consider, for example,
systems that start from a single active site in an otherwise inactive lattice. In this situation,
the central quantity is the survival probability $P_s(t)$ that measures how likely the system is to
be still active (i.e., to contain at least one active site) at time $t$.
For directed percolation problems such as the contact process, the survival probability behaves
in the same way as the density of active sites \cite{GrassbergerdelaTorre79}. Thus, the time-dependencies
(\ref{eq:rho_Griffiths}) and (\ref{eq:rho_clean_critical}) derived for $\rho(t)$ also hold for $P_s(t)$.

We emphasize that the dependencies of the Griffiths dynamical exponent $z'$ on the distance from
criticality given in (\ref{eq:z'}) and (\ref{eq:z'_binary})
hold outside the asymptotic critical region of the disordered contact process.
The analysis of the critical region itself requires more sophisticated methods that will be discussed
in the next section.

\subsection{Critical point}
\label{subsec:Theory_critical}

After discussing the Griffiths phase, we now turn to the critical point of the disordered contact process
itself. The contact process with spatially \emph{uncorrelated} disorder features an exotic infinite-randomness critical point
in the universality class of the (uncorrelated) random transverse-field Ising chain \cite{HooyberghsIgloiVanderzande03,VojtaDickison05}.
Is this critical point stable or unstable against the long-range power-law disorder correlations (\ref{eq:G_power})?
According to Weinrib and Halperin's generalization \cite{WeinribHalperin83} of the Harris criterion,
power-law disorder correlations are irrelevant if the decay exponent $\gamma$ fulfills the inequality
\begin{equation}
\gamma > 2/ \nu_\perp^{unc}
\label{eq:WeinribHalperin}
\end{equation}
where $\nu_\perp^{unc}$ is the correlation length exponent for uncorrelated disorder. If this inequality is violated,
the correlations are relevant, and the critical behavior must change. The correlation length exponent of the contact
process with uncorrelated disorder takes the value $\nu_\perp^{unc}=2$ \cite{Fisher92,HooyberghsIgloiVanderzande03}.
The long-range correlations are thus irrelevant if $\gamma > 1$ and relevant if $\gamma < 1$. Interestingly, this
is the same criterion as we derived for the Griffiths phase in Secs.\ \ref{subsec:Theory_RR} and \ref{subsec:Theory_Griffiths}.

What is the fate of the transition in the long-range correlated case $\gamma < 1$? As long-range correlations tend to further enhance the disorder
effects, we expect the critical behavior to be of infinite-randomness type, but with modified critical exponents that
produce stronger singularities. In the strong-disorder regime close to criticality, the behavior of the contact process is identical
to that of a random transverse-field Ising chain as both are governed by the same strong-disorder renormalization group
recursion relations \cite{Fisher92,HooyberghsIgloiVanderzande03}. Note that the application of these recursion is justified even
in the presence of disorder correlations provided that the distributions of the \emph{logarithms} of $\mu$ and $\lambda$
become infinitely broad. The transverse-field Ising chain with long-range correlated disorder was solved by
Rieger and Igloi \cite{RiegerIgloi99} who mapped the problem onto fractional Brownian motion.
They found an exact result for the tunneling exponent $\psi$ which relates correlation length $\xi_\perp$ and correlation time
$\xi_t$ via $\ln(\xi_t/t_0) \sim \xi_\perp^\psi$. For $\gamma > 1$, it takes the uncorrelated value $\psi=1/2$ while
it is given by $\psi=1-\gamma/2$ for $\gamma < 1$. The correlation length exponent $\nu_\perp$ takes the value 2 for
$\gamma > 1$ as for uncorrelated disorder. For $\gamma < 1$, it reads $\nu_\perp =2/\gamma$ in agreement with general arguments
by Weinrib and Halperin \cite{WeinribHalperin83}. A third exponent is necessary to define a complete set; Rieger and Igloi
numerically calculated the scale dimension $\beta/\nu_\perp$ of the order parameter and found it to decay continuously from its
uncorrelated value $(3-\sqrt{5})/4$ (taken for all $\gamma > 1$) to 0 (for $\gamma=0$).

A qualitative understanding of these results in the context of the contact process can be obtained from simple arguments
based on the strong-disorder
recursion relations \cite{HooyberghsIgloiVanderzande03} even though a closed form solution of the renormalization group
does not exist for the case of long-range correlated disorder
\footnote{This mainly stems from the fact that the strong-disorder renormalization group cannot be formulated in terms of
single-site distributions if the disorder is long-range correlated.}. Imagine performing a (large) number of strong-disorder
renormalization group steps, iteratively removing the largest decay rates $\mu_i$ and infection rates $\lambda_i$.
The resulting chain will consist of surviving sites (representing clusters of original sites)
whose effective decay rate can be estimated as
\begin{equation}
\mu_{\rm eff} = C_\mu \frac {\mu_1 \ldots \mu_L}{\lambda_1 \ldots \lambda_{L-1}} ~
\label{eq:mu_eff}
\end{equation}
 and long bonds with effective infection rates
\begin{equation}
\lambda_{\rm eff} = C_\lambda \frac {\lambda_1 \ldots \lambda_{L}} {\mu_1 \ldots \mu_{L-1}}~
\label{eq:lambda_eff}
\end{equation}
where $L$ is the size of the cluster or bond. In the strong-disorder limit,
the prefactors $C_\mu$ and $C_\lambda$ provide subleading corrections only.
$\ln \mu_{\rm eff}$ and $\ln \lambda_{\rm eff}$ can thus be understood as the displacements of correlated random walks
\begin{equation}
\ln \mu_{\rm eff} \sim \sum_{i=1}^{L-1} \ln(\mu_i/\lambda_i)~, \qquad
\ln \lambda_{\rm eff} \sim \sum_{i=1}^{L-1} \ln(\lambda_i/\mu_i)~.
\label{eq:ln_mu_lambda_eff}
\end{equation}
Right at criticality, these random walks have to be (asymptotically) unbiased because healing
and infection remain competing in the limit $L \to \infty$.
The typical values  $\ln \mu_{\rm typ}$ and $\ln \lambda_{\rm typ}$ of the cluster
healing and infection rates can be estimated from the variance of the random walk displacements
giving
\begin{equation}
|\ln \mu_{\rm typ}| \sim |\ln \lambda_{\rm typ}| \sim \sqrt{L \, \tilde G (L)} \sim \left\{
\begin{array}{cc}
 L^{1/2} & (\gamma >1) \\
 L^{1-\gamma/2} & (\gamma < 1)
\end{array}
\right.
\label{eq:ln_mu_lambda_typ}
\end{equation}
for large $L$.
Here, $\tilde G(L)$ is the sum over the disorder correlation function defined in eq.\ (\ref{eq:tilde_G}).
This estimate thus reproduces the values of $\psi$ quoted above
\footnote{Note that this argument is not rigorous as it neglects the subtle correlations involved in picking
which $\mu_i$ or $\lambda_i$ to decimate in the strong-disorder renormalization group step.}.

Moving away from criticality introduces a bias into the random walks. The crossover from
critical to off-critical behavior occurs when the displacement due to the bias becomes larger
than the displacement (\ref{eq:ln_mu_lambda_typ}) due to the randomness.
The bias term scales as $|\lambda-\lambda_c|L$. We thus obtain a crossover length
\begin{equation}
L_x \sim \left\{
\begin{array}{cc}
 |\lambda-\lambda_c|^{-2} & (\gamma >1) \\
 |\lambda-\lambda_c|^{-2/\gamma} & (\gamma < 1)
\end{array}
\right.
\label{eq:crossover_L}
\end{equation}
in agreement with the quoted values of $\nu_\perp$.

\section{Monte-Carlo simulations}
\label{sec:MC}
\subsection{Overview}
\label{subsec:MC_Overview}

We now turn to large-scale Monte-Carlo simulations of the one-dimensional
contact process with power-law correlated disorder. We use the same numerical implementation
of the contact process as in earlier studies  with uncorrelated
disorder in one, two, three, and five dimensions in Refs.\ \cite{VojtaDickison05,VojtaFarquharMast09,Vojta12,VojtaIgoHoyos14}.
It is based on an algorithm suggested by Dickman \cite{Dickman99}:
The simulation starts at time $t=0$ from
an initial configuration of active and inactive sites and consists of a sequence of events. During each event
an active site $i$ is chosen at random from a list of all $N_a$ active sites. Then a process is selected,
either infection of a neighbor with probability $\lambda_i/(1+ \lambda_i)$ or healing with probability $1/(1+ \lambda_i)$.
For infection, either the left or the right neighbor are chosen with probability 1/2. The infection succeeds if this neighbor
is inactive. The time is then incremented by $1/N_a$.

Using this algorithm, we have simulated long chains for times up to $t=10^7$.
All production runs use $L=2^{20} \approx 10^6$ sites with periodic boundary conditions,
and the results are
averages over large numbers of disorder configurations; precise data will be given below.

The random infection rates $\lambda_i$ are drawn from  a correlated binary distribution in which
$\lambda_i$ can take values $\lambda$ and $c\lambda$ with overall probabilities $(1-p)$ and $p$, respectively.
Here, $p$ and $c$ are constants between 0 and 1. To generate these correlated random variables, we employ
the Fou\-rier-filtering method \cite{MHSS96}. It starts from uncorrelated Gaussian random numbers $u_i$ and turns them into correlated Gaussian
random numbers $v_i$ characterized by the (translationally invariant) correlation function $G_\lambda(i,j)$.
This is achieved by transforming the Fourier components $\tilde{u}_q$ of the uncorrelated random numbers according to
\begin{equation}
\label{ffm_def}
{\tilde{v}_q} = \left[ \tilde G(L,q) \right]^{1/2} \tilde{u}_q,
\end{equation}
where $\tilde G(L,q)$ is the Fourier transform of $G_\lambda(i,j)$.
We parameterize our long-range correlations by the function
\begin{equation}
G_\lambda(i,j) = \left [1+(i-j)^2 \right]^{-\gamma/2}
\label{eq:G_numerics}
\end{equation}
with periodic boundary conditions using the minimum image convention. Simulations are performed for
$\gamma=1.5, 0.8, 0.6$ and $0.4$.
To arrive at binary random variables, the correlated Gaussian random numbers $v_i$ then undergo binary projection: the infection rate $\lambda_i$
takes the value $\lambda$ (``strong site'') if $v_i$ is greater than a composition-dependent threshold and the value $c\lambda$ with $0<c<1$ (``weak site'') if $v_i$ is less than the threshold. We chose a concentration $p=0.8$ of weak sites and a strength $c=0.2$ in all simulations.
While the binary projection changes the details of the disorder correlations, the functional form of the long-distance tail remains unchanged.

Most of our simulations are spreading runs that start from a single active site in an otherwise inactive
lattice; we monitor the survival probability $P_s(t)$, the number of sites $N_s(t)$
of the active cluster, and its (mean-square) radius $R(t)$. Within the activated scaling scenario \cite{HooyberghsIgloiVanderzande03,VojtaDickison05}
associated with an infinite-randomness critical point, these quantities are expected
to display logarithmic time dependencies,
\begin{eqnarray}
P_s &\sim& [\ln(t/t_0)]^{-\bar\delta}~,
\label{eq:P_s(t)}\\
N_s &\sim& [\ln(t/t_0)]^{\bar\Theta}~,
\label{eq:N_s(t)}\\
R &\sim& [\ln(t/t_0)]^{1/\psi}~.
\label{eq:R(t)}
\end{eqnarray}
The exponents $\bar\delta$ and $\bar\Theta$ can be expressed in terms of the scale dimension $\beta/\nu_\perp$
of the order parameter and the tunneling exponent $\psi$ as
$\bar\delta = \beta/(\nu_\perp\psi)$ and $\bar\Theta= 1/\psi - 2\bar\delta$ \cite{VojtaDickison05}.

\subsection{Results: critical behavior}
\label{subsec:MC_critical}

We start be considering the case $\gamma=1.5$. According to the theory laid out in Sec.\ \ref{sec:Theory},
the power-law disorder correlations are irrelevant for $\gamma > 1$. We therefore expect the critical behavior
for $\gamma=1.5$ to be identical to that of the random contact process with uncorrelated disorder
which features an infinite-randomness critical point in the universality class of the (uncorrelated) random transverse-field
Ising chain \cite{HooyberghsIgloiVanderzande03,VojtaDickison05}. Its critical exponents are known exactly, their numerical values read
$\beta=0.38197$, $\nu_\perp=2$, $\psi=0.5$,
$\bar\delta=0.38197$, and $\bar\Theta=1.2360$ \cite{Fisher92,Fisher95}.

\begin{figure}
\includegraphics[width=8.5cm]{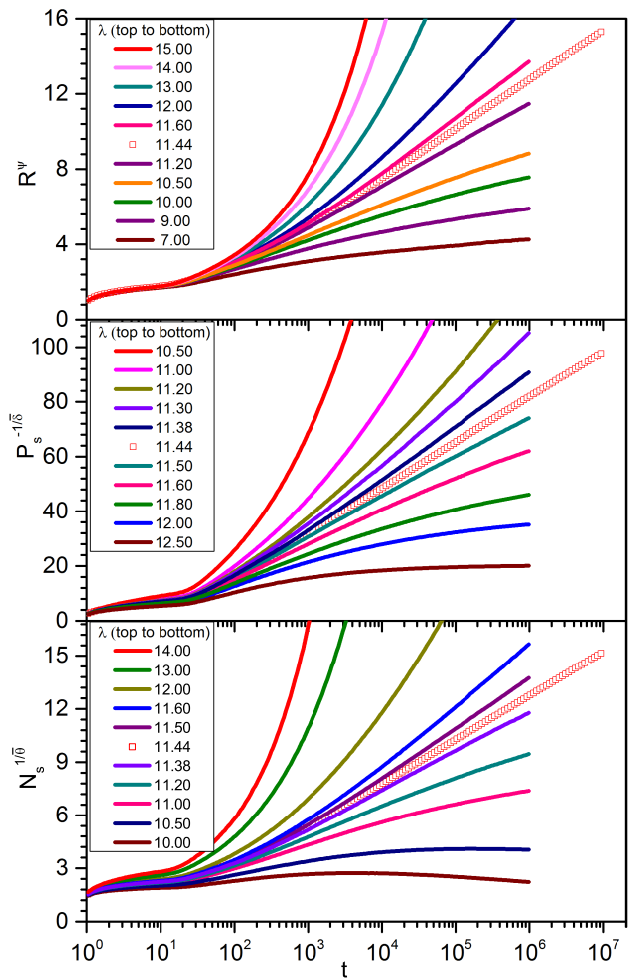}
\caption{(Color online) Time evolution of the number of active sites $N_s$, the survival probability $P_s$, and
the radius of the active cloud $R$ for the disordered contact process with power-law disorder correlations
characterized by a decay exponent $\gamma=1.5$. The data are averages over up to $40000$ samples
 with $100$ individual runs per sample.
The critical exponents are fixed at their uncorrelated values $\psi=0.5$,
$\bar\delta=0.38197$, and $\bar\Theta=1.2360$.}
\label{fig:Fig_01}
\end{figure}

To test these predictions, we analyze the time evolution of $P_s$, $N_s$ and $R$ in Fig.\ \ref{fig:Fig_01}.
Specifically, the figure presents plots of $P_s^{-1/\bar\delta}$, $N_s^{1/\bar\Theta}$ and $R^{\psi}$
vs.\ $\ln(t)$ using the theoretically predicted exponent values. In such plots, the critical time dependencies
(\ref{eq:P_s(t)}) to (\ref{eq:R(t)}) correspond to straight lines independent of the unknown value of the
microscopic time scale $t_0$.   The plots show that the data for infection rate $\lambda=11.44$
follow the predicted time dependencies (\ref{eq:P_s(t)}) to (\ref{eq:R(t)}) over more than four orders of
magnitude in time. We thus identify $\lambda_c=11.44(6)$ as the critical infection rate (the number in
brackets is an estimate of the error of the last digit);
and we conclude that the critical behavior for $\gamma=1.5$ is indeed identical to that of the
contact process with uncorrelated disorder.

\begin{figure}
\includegraphics[width=8.5cm]{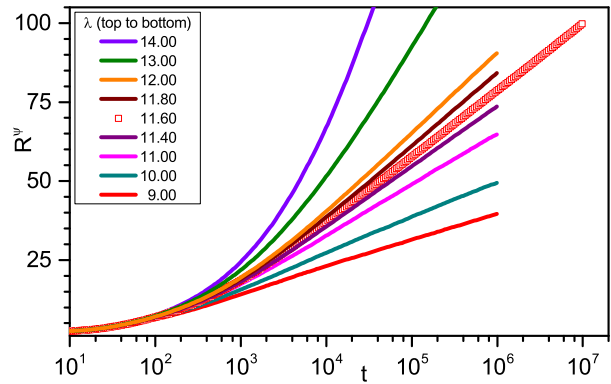}
\caption{(Color online) Time evolution of the radius of the active cloud $R$ for
$\gamma=0.4$. The data are averages over about $30000$ samples with $100$ individual runs per sample.
The tunneling exponent is set to its analytical value $\psi=1-\gamma/2=0.8$.}
\label{fig:Fig_015}
\end{figure}

We now turn to $\gamma < 1$, for which the long-range correlations are expected to change the critical behavior.
A complete set of exponents is not known analytically in this case; the data analysis is therefore more complicated
than for $\gamma > 1$.
As we do have an analytical value for the tunneling exponent, $\psi=1-\gamma/2$, we can graph $R^{\psi}$ vs.\ $\ln(t)$,
to find the critical point. Figure \ref{fig:Fig_015} shows the corresponding plot for $\gamma=0.4$.
The data at $\lambda=11.6$ follow the predicted time dependence (\ref{eq:R(t)}) for more than
three orders of magnitude in time. We thus identify  $\lambda_c=11.6(2)$
as the critical infection rate.
Analogous plots for $\gamma=0.8$ and $0.6$ give infection rates of $\lambda_c=11.3(2)$
and $\lambda_c=11.4(2)$, respectively.

\begin{figure}
\includegraphics[width=8.5cm]{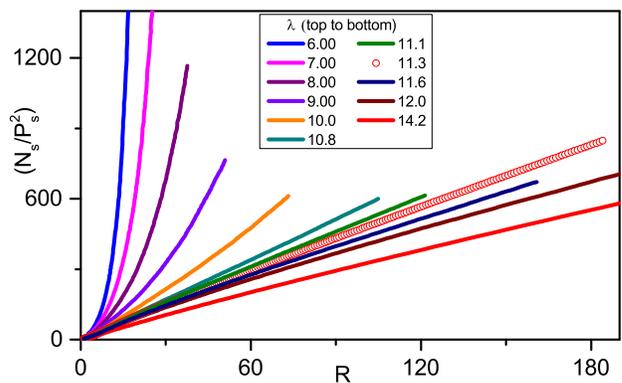}
\caption{(Color online) $N_s/P_s^2$ vs.\ $R$ for a correlation decay exponent $\gamma=0.8$.
 The data are averages over about $20000$ samples
 with $100$ individual runs per sample.
The maximum time is $10^6$ for all curves except the critical one, $\lambda=11.3$, for which
it is $10^7$. }
\label{fig:Fig_02}
\end{figure}

Alternatively, we can employ a version of the method used in Refs.\  \cite{VojtaFarquharMast09,Vojta12}
that allows us to eliminate the unknown microscopic time scale $t_0$ from the analysis.
It is based on the observation that $t_0$ takes the same value in all of the quantities
(because it is related to the basic energy scale of the underlying renormalization group).
Thus, if we plot $N_s(t)$ versus $P_s(t)$, the critical point corresponds to power-law
behavior, and $t_0$ drops out. The same is true for other combinations of observables.
Specifically, by combining eqs.\ (\ref{eq:P_s(t)}), (\ref{eq:N_s(t)}) and (\ref{eq:R(t)}), we see that
$N_s/P_s^2 \propto R$ at criticality. Thus, identifying straight lines in plots of $N_s/P_s^2$ versus $R$
allows us to find the critical point without needing a value for  $t_0$.
Figure \ref{fig:Fig_02} shows such a plot for $\gamma=0.8$;
and we have created analogous plots of $\gamma=0.6$ and $0.4$.
They give the same critical infection rates, $\lambda_c=11.3(2)$ (for $\gamma=0.8$), $\lambda_c=11.4(2)$ (for $\gamma=0.6$),
and $\lambda_c=11.6(2)$ (for $\gamma=0.4$) as the plots of $R^\psi$ vs.\ $\ln(t)$.
Interestingly, within their numerical errors, $\lambda_c$ does not depend on the decay exponent $\gamma$ of the disorder correlations.

\begin{figure}
\includegraphics[width=8.5cm]{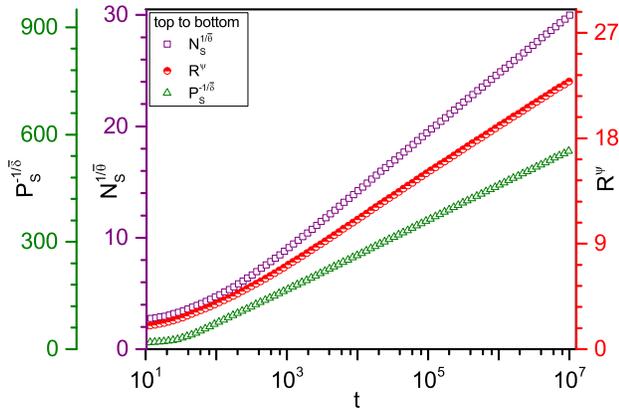}
\caption{(Color online) $N_s^{1/\bar\Theta}$, $P_s^{-1/\bar\delta}$, and $R^{\psi}$ versus $\ln(t)$ at criticality
 for a correlation decay exponent $\gamma=0.8$. Here, $\psi=0.6$ is set to its theoretical value
 while $\bar\delta=0.269$ and $\bar\Theta=0.982$ are
determined from the data by requiring that the corresponding curves become straight lines for large times.}
\label{fig:Fig_05}
\end{figure}

Once the critical point is identified, we can verify and/or find critical exponents by analyzing the time evolutions of $P_s$, $N_s$ and $R$.
 Figure \ref{fig:Fig_05}
displays $P_s^{-1/\bar\delta}$, $N_s^{1/\bar\Theta}$ and $R^{\psi}$ versus $\ln(t)$ at criticality for $\gamma=0.8$.
The tunneling exponent $\psi$ is set to its theoretical value $1-\gamma/2$ while $\bar\delta$ and $\bar\Theta$ are
determined from the data by requiring that the corresponding curves become straight lines for large times.
The data follow the predicted logarithmic time dependencies (\ref{eq:P_s(t)}), (\ref{eq:N_s(t)}) and (\ref{eq:R(t)})
over about four orders of magnitude in time. This not only confirms the theoretical value of $\psi$,
it also allows us to extract estimates the scale dimension $\beta/\nu_\perp$ of the order parameter from both $\bar\delta$
and $\bar\Theta$. We have performed the same analysis also for $\gamma=0.6$ and $\gamma=0.4$.

The resulting exponent values are summarized in Table \ref{tab:exponents}.
\begin{table}[tb]
\renewcommand*{\arraystretch}{1.2}
\begin{tabular*}{8cm}{@{\extracolsep{\fill}}c|cccc}
\hline\hline
exponent         & $\gamma > 1$ & $\gamma=0.8$ & $\gamma=0.6$ & $\gamma=0.4$ \\
\hline
 $\nu_\perp$     &     2         &    2.5       &    3.33      &     5       \\
 $\psi$          &     0.5       &    0.6       &    0.7       &     0.8     \\
\hline
 $\bar\delta$     & 0.3820       &  0.27        &   0.20       &    0.13  \\
 $\bar\Theta$     & 1.2360       &  0.98        &   0.98       &    1.01 \\
 $\beta/\nu_\perp$& 0.1910       &  0.18        &   0.14       &    0.10   \\
\hline\hline
\end{tabular*}
\caption{Critical exponents of the one-dimensional contact process with power-law
correlated disorder. The exponents $\nu_\perp$ and $\psi$ (above the horizontal line)
are known analytically, as are all exponents in the short-range case $\gamma > 1$.
The exponents $\bar\delta$ and $\bar\Theta$ for $\gamma <1$ stem from  fits of
our data. The scale dimension $\beta/\nu_\perp$ of the order parameter can be extracted
from both $\bar\delta$ and $\bar\Theta$, the data in the table are averages of the two values.}
\label{tab:exponents}
\end{table}
The uncertainty of  $\bar\delta$ and $\bar\Theta$ can be roughly estimated from
the hyperscaling relation $\bar\Theta+2\bar\delta=1/\psi$.
The exponents for $\gamma=0.6$ and $\gamma=0.4$ fulfill this relation in good approximation
(less than 4\% difference between the left and the right sides). For $\gamma=0.8$, the agreement
is not quite as good. As $\gamma=0.8$ is close to the marginal value of 1, this may be caused by a slow crossover
from the short-range correlated fixed point to the long-range correlated one.

The values of the scale dimension of the order parameter, $\beta/\nu_\perp$,
are in reasonable agreement with those
calculated by Rieger and Igloi from the average persistence of a Sinai
random walker (see inset of Fig.\ 1 of  Ref.\ \cite{RiegerIgloi99}).

\begin{figure}
\includegraphics[width=8.5cm]{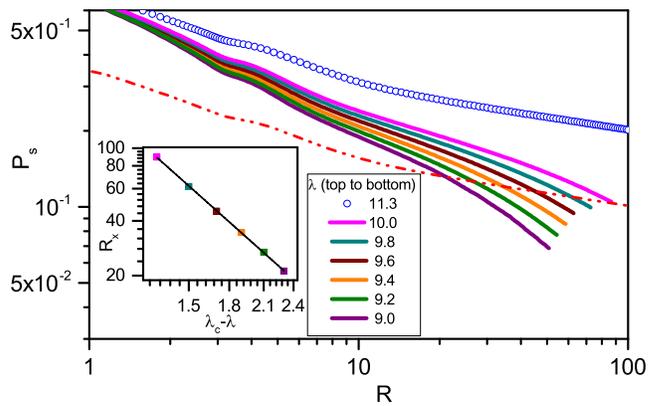}
\caption{(Color online) Double-log plot of $P_s$ vs.\ $R$ for decay exponent
$\gamma=0.8$ and several infection rates $\lambda$ at and below the critical
rate $\lambda_c=11.3$. The dash-dotted line shows $P_s/2$ for $\lambda=\lambda_c$.
The crossing points of the dash-dotted line with the off-critical data determines
the crossover radius $R_x$. Inset: $R_x$ vs.\ $|\lambda-\lambda_c|$. The solid line
is a power-law fit to $R_x \sim |\lambda-\lambda_c|^{-\nu_\perp}$ with an exponent
$\nu_\perp=2.5$. }
\label{fig:Fig_018}
\end{figure}

To obtain a complete set of exponents, we also analyze off-critical data.
Fig.\ \ref{fig:Fig_018} shows a double-logarithmic plot of $P_s$ vs.\ $R$
for decay exponent $\gamma=0.8$.
The plot allows us to determine the crossover
radius $R_x$ at which the survival probability of slightly off-critical curves
has dropped to half of its critical value. According to scaling, the crossover radius
must depend on the distance from criticality via $R_x \sim |\lambda-\lambda_c|^{-\nu_\perp}$.
The inset of Fig.\ \ref{fig:Fig_018} shows that our data indeed follow
this power law with the predicted exponent $\nu_\perp=2/\gamma=2.5$.

\subsection{Results: Griffiths phase}
\label{subsec:MC_Griffiths}

We now turn to the Griffiths phase $\lambda_{c0} \le \lambda < \lambda_c$
where $\lambda_{c0}\approx 3.298$ is the critical infection rate of the clean
contact process containing only ``strong'' sites ($p=0$).

\begin{figure}
\includegraphics[width=8.5cm]{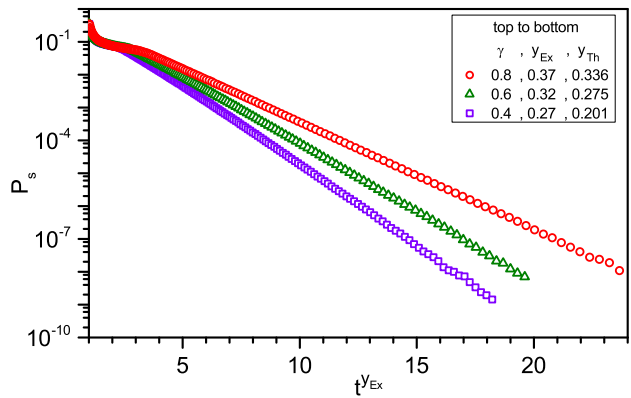}
\caption{(Color online) Time evolution of the survival probability $P_s$ at the clean critical infection rate
$\lambda_{c0}=3.298$ for decay exponents $\gamma=0.8$, 0.6, and 0.4. The data are averages over $2\times 10^4$ to $10^5$ samples
 with at least $10^4$ individual runs per sample. The experimental values $y_{Ex}$ are determined by requiring that the respective
 curves become straight lines for large times, implying a stretched exponential time dependence,  $\ln P_s \sim t^y$.
 The theoretical values follow from eq.\ (\ref{eq:rho_clean_critical}) which gives $y_{Th}=\gamma/(\gamma +z_0)$.}
\label{fig:Fig_08}
\end{figure}

Right at the clean critical point, $\lambda=\lambda_{c0}$, the time evolution of the survival probability
is predicted to follow the stretched exponential (\ref{eq:rho_clean_critical})
in the long-time limit.  Our corresponding data for $\gamma=0.8$, 0.6, and 0.4
are plotted in Fig.\ \ref{fig:Fig_08}.
For all $\gamma$, the data indeed follow stretched exponentials over more than six orders of magnitude in $P_s$.
The exponent $y$ decreases with decreasing $\gamma$, as predicted in (\ref{eq:rho_clean_critical}).
The actual numerical values of $y$ are somewhat larger than the prediction $y=\gamma/(\gamma +z_0)$.
We attribute this to the fact that, due to the rapid decay of $P_s$, the data are taken at rather short times ($t\lessapprox 10^3$).
Thus, they probably have not reached the true asymptotic regime, yet.

\begin{figure}
\includegraphics[width=8.5cm]{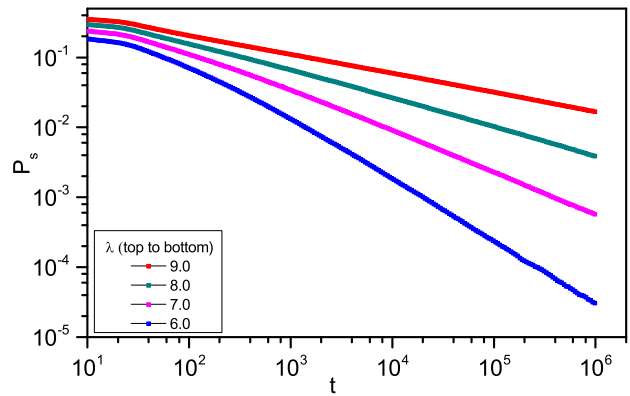}
\caption{(Color online) Double-log plot of the survival probability $P_s$ vs.\ time $t$ for decay
exponent $\gamma=1.5$ at several infection rates
inside the Griffiths phase, $\lambda_{c0} < \lambda < \lambda_c$.
The data are averages over up to $40000$ samples  with $100$ individual runs per sample.}
\label{fig:Fig_016}
\end{figure}

We now move into the bulk of the Griffiths phase, $\lambda_{c0} < \lambda < \lambda_c$.
Here, we wish to contrast the conventional power-law Griffiths singularity with the
unusual non-power-law form (\ref{eq:rho_Griffiths}).
Fig.\ \ref{fig:Fig_016} shows the survival probability as a function of time for a decay
exponent $\gamma=1.5$ and several infection rates inside the Griffiths phase.
After initial transients, all data follow power laws (represented by straight lines) over serval
orders of magnitude in
$P_s$ and/or $t$. For $\gamma=1.5$, we thus find the same type of power-law Griffiths singularity
as in the case of uncorrelated or short-range correlated disorder.

\begin{figure}
\includegraphics[width=8.5cm]{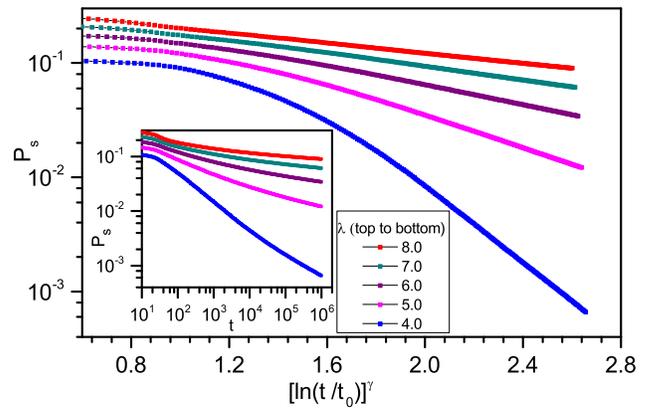}
\caption{(Color online) Survival probability $P_s$ vs.\ time $t$ for decay exponent $\gamma=0.4$ at several infection rates
inside the Griffiths phase, $\lambda_{c0} < \lambda < \lambda_c$, plotted such that eq.\ (\ref{eq:rho_Griffiths})
yields straight lines (the values of $t_0$ are fit parameters).
The data are averages over up to $20000$ samples  with $100$ individual runs per sample.
Inset: Double-log plot of the same data to test for power-law behavior.}
\label{fig:Fig_011}
\end{figure}

In the long-range correlated regime, $\gamma < 1$, we expect the survival probability to follow eq.\
(\ref{eq:rho_Griffiths}) rather than a power-law.
This prediction is tested in Fig.\ \ref{fig:Fig_011} which shows of $P_s$ vs.\ $t$ for decay
exponent $\gamma=0.4$.
In the double-logarithmic plot in the inset, all data show pronounced upward curvatures rather than the straight lines expected for power laws.
In contrast, when plotted as $\ln P_s$ vs.\ $[\ln (t/t_0)]^\gamma$ (where $t_0$ is a fit parameter)
in the main panel of the figure, all curves become straight
for sufficiently long times implying that the long-time behavior of $P_s$ indeed follows eq.\ (\ref{eq:rho_Griffiths}).

We have produced analogous plots for decay exponents $\gamma=0.6$ and 0.8. As $\gamma$ decreases from 1 towards 0, the upward curvature
in the double-logarithmic plots becomes bigger, reflecting stronger and stronger deviations from power-law behavior,
as expected. In contrast, eq.\ (\ref{eq:rho_Griffiths}) describes the long-time behavior of all data very well, confirming our
theory.

\section{Generalizations}
\label{sec:General}
\subsection{Higher dimensions}

It this section, we generalize our results to the contact process in higher dimensions $d > 1$.  The theory of
Sec.\ \ref{subsec:Theory_RR} can be easily adapted, yielding the rare-region distribution
\begin{equation}
P(\lambda_{RR},L_{RR}) \sim \left\{
\begin{array}{cc} \exp \left[ - \dfrac 1 {2b^2} \, {L_{RR}^d}\,      (\lambda_{RR}-\bar\lambda)^2  \right] & (\gamma > d) \\
                  \exp \left[ - \dfrac 1 {2b^2} \, {L_{RR}^\gamma}\, (\lambda_{RR}-\bar\lambda)^2  \right] & (\gamma < d)
\end{array}
\right. ~.
\label{eq:PlambdaRR-general-d}
\end{equation}
This means that the functional form of the rare-region distribution is identical to the case of uncorrelated disorder
as long as $\gamma > d$. For $\gamma < d$, the probability for finding a rare-region decays more slowly with its size.
In terms of the volume $L_{RR}^d$, it is given by a stretched exponential rather than a simple one.

Using this result, we now repeat the calculation of Sec.\ \ref{subsec:Theory_Griffiths} for general $d$. For $\gamma < d$, the resulting
long-time behavior of the density of active sites in the Griffiths phase reads
\begin{equation}
\rho(t) \sim \exp\left[ -\frac d {z'} \left(\ln \frac t{t_0} \right)^{\gamma/d} \right]
\label{eq:rho_Griffiths-general-d}
\end{equation}
with
\begin{equation}
z' \sim d (\lambda_c - \bar\lambda)^{\gamma\nu_{0\perp}-2}
\label{eq:z'-general-d}
\end{equation}
As in one dimension, the decay described by eq.\ (\ref{eq:rho_Griffiths-general-d})
is slower than any power. For $\gamma > d$, in contrast, we find the usual power-law behavior.
Equation (\ref{eq:rho_Griffiths-general-d}) also holds for a correlated binary distribution
with $z' = d\, a^{\gamma/d} /c$. The behavior right at the boundary of the Griffiths phase
(when the stronger of the two infection rates of the binary distribution is tuned to the clean
critical value) takes the form (\ref{eq:rho_clean_critical}) for all dimensions.

The behavior of the critical point itself will again be of infinite-randomness type, but
for a sufficiently small correlation decay exponent $\gamma < 2/\nu_\perp^{unc}$, the critical exponents will differ
from those of the contact process with uncorrelated disorder (which were found numerically in Ref.\
\cite{VojtaFarquharMast09} for two dimensions and in Ref.\ \cite{Vojta12} for three dimensions).
The correlation length exponent will take the value $\nu_\perp=2/\gamma$ \cite{WeinribHalperin83};
other exponents need to be found numerically \cite{RiegerIgloi99}.

It is interesting to compare the relevance criteria of the long-range correlations in the Griffiths phase
and at the critical point. In one dimension,  the long-range correlations become relevant for $\gamma < 1$
both in the Griffiths phase and at criticality (because the correlation length exponent
of the one-dimensional contact process with uncorrelated disorder has the value $\nu_\perp^{unc} =2$, saturating
the Harris criterion). In dimensions $d>1$, the two criteria differ. The uncorrelated correlation length exponent is larger than
$2/d$ \cite{VojtaFarquharMast09,Vojta12}. Thus, the long-range correlations do not become
relevant for $\gamma < d$ but only if $\gamma < 2/\nu_\perp^{unc} < d$. In contrast, the
long-range correlations become relevant for $\gamma < d$ in the Griffiths phase.
Consequently, for $d>1$, we expect a (narrow) range of decay exponents $\gamma$ for which the long-range correlations
are relevant in the Griffiths phase but irrelevant at criticality. The fate of the system in this
subtle regime remains a task for the future.

\subsection{Other systems}
\label{subsec:Other_systems}

The theory of Secs.\ \ref{subsec:Theory_RR} and \ref{subsec:Theory_Griffiths} and its
generalization to higher dimensions have produced enhanced \emph{non-power-law} Griffiths
singularities for sufficiently long-ranged disorder correlations. Are these results restricted
to the contact process or do they apply to other systems as well? In this section, we show that
they hold for a broad class of systems in which the charateristic energy or inverse time scale
of a rare region depends exponentially on its volume (class B of the rare region classification
of Refs.\ \cite{Vojta06,VojtaSchmalian05}). In addition to the contact process, this class
contains, e.g., the random transverse-field Ising model, Hertz' model of the itinerant antiferromagnetic
quantum phase transition, and the pair-breaking superconductor-metal quantum phase transition.

To demonstrate the enhanced Griffiths singularities, we generalize the calculation of the
rare region density of states developed in Ref.\ \cite{VojtaHoyos14} to the case of our
power-law correlated disorder. Consider a disordered system with rare regions whose
characteristic energy $\epsilon$ depends on their volume via
\begin{equation}
\epsilon(\lambda_{RR},L_{RR})=\epsilon_0 \exp[-a L_{RR}^d]~.
\label{eq:epsilon}
\end{equation}
Here, $\epsilon_0$ is a microscopic energy scale, and $a=a'(\lambda_{RR}-\lambda_c)^{d\nu_{0\perp}}$ with $\lambda$ representing the parameter
that tunes the system through the phase transition.
In the contact process, $\epsilon=1/\tau$ is the inverse life time of a rare region;
in the transverse-field Ising model, it represents its energy gap.
We can derive a rare-region density of states by summing over all values of $\lambda_{RR}$ and $L_{RR}$,
\begin{equation}
\tilde\rho(\epsilon) \sim \int_{\lambda_c}^{\infty} \int_{0}^{\infty} dL_{RR}\,  P(\lambda_{RR},L_{RR}) \, \delta[\epsilon-\epsilon(\lambda_{RR},L_{RR})]
\label{eq:DOS}
\end{equation}
with the Gaussian rare-region probability $P(\lambda_{RR},L_{RR})$ from eq.\ (\ref{eq:PlambdaRR-general-d}).
After carrying out the integral over $L_{RR}$ with the help of the $\delta$ function, the remaining
$\lambda_{RR}$-integral can be performed in saddle-point approximation in the limit $\epsilon \to 0$.
For $\gamma < d$, the resulting density of states takes the form
\begin{equation}
\tilde\rho(\epsilon) \sim \frac 1 \epsilon \exp\left[ -\frac d {z'} \left(\ln \frac {\epsilon_0}\epsilon \right)^{\gamma/d} \right]
\label{eq:DOS_Griffiths-general-d}
\end{equation}
with $z'$ given by eq.\ (\ref{eq:z'-general-d}). For $\gamma > d$, in contrast, we recover the usual power-law behavior
$\tilde\rho(\epsilon) \sim \epsilon^{d/z'-1}$.
If we start from correlated binary disorder rather than a Gaussian distribution, we arrive at the same
expression (\ref{eq:DOS_Griffiths-general-d}) for the density of states with $z'= d a^{\gamma/d} /c$.
Equation (\ref{eq:DOS_Griffiths-general-d}) shows that the Griffiths singularities are qualitatively enhanced
for $\gamma < d$ as the density of states diverges as $1/\epsilon$ times a function that is slower than any
power law.

Griffiths singularities in other observables can be calculated from appropriate integrals of
$\tilde\rho(\epsilon)$. For example, our results for the density of active sites in the contact process
can be reproduced by $\rho(t) \sim  \int d\epsilon\,\tilde\rho(\epsilon) \exp(-\epsilon t)$. In the case of the
random transverse-field Ising model, we can calculate (see, e.g., Ref.\ \cite{Vojta10}) the temperature dependence of observables
such as the entropy $S(T)\sim \int_0^T d\epsilon \tilde\rho(\epsilon)$, the specific heat $C(T)=T(\partial S/\partial T)$,
and the susceptibility $\chi(T) \sim (1/T)\int_0^T d\epsilon \tilde\rho(\epsilon)$. For $\gamma < d$, we find
\begin{equation}
 S(T) \sim C(T) \sim T\chi(T) \sim \exp\left[ -\frac d {z'} \left(\ln \frac {\epsilon_0} T \right)^{\gamma/d} \right]~.
\label{eq:S(T)}
\end{equation}
Analogously, the magnetization in a longitudinal field $H$ scales as
\begin{equation}
M(H) \sim \exp\left[ -\frac d {z'} \left(\ln \frac {\epsilon_0} H \right)^{\gamma/d} \right]~.
\label{eq:M(H)}
\end{equation}
Let us compare these results with those obtained in Ref.\ \cite{RiegerIgloi99}.
Equations (\ref{eq:DOS_Griffiths-general-d}), (\ref{eq:S(T)}), and (\ref{eq:M(H)}) yield Griffiths
singularities that are qualitatively stronger than power laws. In contrast, Rieger and Igloi obtained
the usual power-law Griffiths singularities, albeit with changed exponents. We believe that this discrepancy arises from the fact that
Rieger and Igloi assumed that the probability for finding a strongly coupled cluster of size $L_{RR}$ in $d$ dimensions
takes the same functional form, $\exp(-c L_{RR}^d)$, as for uncorrelated disorder. Our calculations show that this assumption is
justified for $\gamma > d$. For $\gamma < d$, however, the rare region probability decays as $\exp(-c L_{RR}^\gamma)$,
i.e., more slowly than in the uncorrelated case.

\section{Conclusions}
\label{sec:Conclusions}

To summarize, we have studied the effects of long-range spatial disorder correlations on the critical behavior and
the Griffiths singularities in the disordered one-dimensional contact process.  As long as the correlations decay
faster as $1/r_{ij}$ with the distance $r_{ij}$ between the sites, the correlations are irrelevant both
at criticality and in the Griffiths phase. This means that both the critical and the Griffiths singularities are
identical to those of the contact process with uncorrelated disorder. If the correlations decay more slowly than
$1/r_{ij}$, the universality class of the critical point changes, and the Griffiths singularities take an
enhanced, non-power-law form.

What is the reason for the enhanced singularities? As positive spatial correlations imply that neighboring sites have similar
infection rates, it is intuitively clear that sufficiently long-ranged correlations must increase the probability for finding
large atypical regions. This is borne out in our calculations in Sec.\   \ref{subsec:Theory_RR}: If the disorder correlations
decay more slowly than $1/r_{ij}$, the probability for finding a rare region behaves as a stretched exponential of its size
(rather than the simple exponential found for uncorrelated and short-range correlated disorder). Note that similar stretched
exponentials have also been found in the distributions of rare events in long-range correlated time series
\cite{BEKH05,AltmanKantz05}.

Our theory of the Griffiths phase is easily generalized to higher dimensions. In general dimension $d$, the rare-region
probability decays exponentially with the rare-region volume  as long as the disorder correlations decay faster than  $1/r_{ij}^d$.
As a result, the Griffiths
singularities take the usual power-law form. For correlations decaying slower than
  $1/r_{ij}^d$, the rare region probability becomes a stretched exponential
of the volume, leading to enhanced, non-power-law Griffiths singularities.

Moreover, as shown in Sec.\ \ref{subsec:Other_systems}, the theory is not restricted to the contact process. It holds for
all systems for which the characteristic energy (or inverse time) of a rare region depends exponentially on its volume,
i.e., for all systems in class B of the rare region classification
of Refs.\ \cite{Vojta06,VojtaSchmalian05}. The random transverse-field Ising model is a prototypical example in the class.
Our theory predicts that the character of its Griffiths singularities changes from the usual power-law behavior for correlations
decaying faster than  $1/r_{ij}^d$ to the enhanced non-power-law forms  (\ref{eq:S(T)}) and  (\ref{eq:M(H)})
for correlations decaying slower than  $1/r_{ij}^d$.

What about systems in the other classes, class A and class C, of the rare region classification of Refs.\
\cite{Vojta06,VojtaSchmalian05}?
The rare regions in systems belonging to class A have characteristic energies that decrease
as a power of their sizes. Using this power-law dependence rather than the exponential
(\ref{eq:epsilon}) in the calculation of Sec.\ \ref{subsec:Other_systems} yields
an exponentially small density of states. We conclude that rare regions effects in class A
remain very weak, even in the presence of long-range disorder correlations.
Rare regions in systems belonging to class C can undergo the phase transition by themselves,
independently from the bulk system. This results in a smearing of the global phase transition.
Svoboda et al.\ \cite{SNHV12} considered the effects of spatial disorder correlations
on such smeared phase transitions. They found that even
short-range correlations can have dramatic effects and qualitatively change the behavior
of observable quantities compared to the uncorrelated case. This phenomenon may have
been observed in Sr$_{1-x}$Ca$_x$RuO$_3$ \cite{Demkoetal12}.

\section*{Acknowledgements}

This work was supported by the NSF under Grant Nos.\ DMR-1205803
and PHYS-1066293. We acknowledge the hospitality of the Aspen Center for Physics.

\bibliographystyle{apsrev4-1}
\bibliography{../../00Bibtex/rareregions}
\end{document}